\documentstyle[preprint,aps,multirow,amssymb]{revtex}

\tightenlines

\begin{document}
\draft

\title{Structural Properties of Self-Attracting Walks}

\author{A.~Ordemann$^{1,2}$, E.~Tomer$^{2}$, G.~Berkolaiko$^{3}$, 
        S.~Havlin$^{2}$, and~A.~Bunde$^{1}$}

\address{$^1$Institut~f\"ur~Theoretische~Physik~III,
         Justus-Liebig-Universit\"at~Giessen,
         Heinrich-Buff-Ring~16,   
         35392~Giessen, Germany}
        
\address{$^2$Minerva~Center~and~Department~of~Physics,
         Bar-Ilan~University,
         52900~Ramat-Gan, Israel}

\address{$^3$Department~of~Physics~of~Complex~Systems,
         Weizmann~Institute~of~Science,
	     76100~Rehovot, Israel}
         
\date{\today}

\maketitle
         
\begin{abstract} 

Self-attracting walks ({\textit{SATW}}) with attractive interaction $u > 0$
display a swelling-collapse transition at a critical  $u_{\mathrm{c}}$ for
dimensions $d \ge 2$, analogous to the $\Theta$ transition of polymers.  We are
interested in the structure of the clusters generated by {\textit{SATW}} below
$u_{\mathrm{c}}$ (swollen walk), above  $u_{\mathrm{c}}$ (collapsed walk), and
at $u_{\mathrm{c}}$, which can be characterized by the fractal dimensions of
the clusters $d_{\mathrm{f}}$ and their interface $d_{\mathrm{I}}$. Using
scaling arguments and Monte Carlo simulations, we find that for
$u<u_{\mathrm{c}}$,  the structures are in the universality class of clusters
generated by simple random walks. For $u>u_{\mathrm{c}}$, the clusters are
compact, i.e.~$d_{\mathrm{f}}=d$ and $d_{\mathrm{I}}=d-1$. At $u_{\mathrm{c}}$,
the {\textit{SATW}} is in a new universality class. The clusters are  compact
in both $d=2$ and $d=3$, but their interface is fractal: $d_{\mathrm{I}}=1.50
\pm 0.01$ and $ 2.73\pm 0.03$  in $d=2$ and $d=3$, respectively.  In $d=1$,
where the walk is collapsed for all $u$ and no swelling-collapse transition
exists,  we derive analytical expressions for the average number of visited
sites  $\langle S\rangle$ and the mean time $\langle t\rangle$ to visit $S$
sites. 

\end{abstract} 


\newpage
\section{Introduction}

In previous years various models of random walks (\textit{RW}) with memory or
interaction have been
studied~\cite{Stanley83,Redner83,Chan84,Duxbury85,Domb72,Domb83,Oettinger85,Amit83,Sapoz94,Malakis75},
in order to account for distinct features of physical, chemical, and
biological  systems whose complexity goes beyond what can be obtained from the
simple random walk picture. Perhaps the most extensive studied model is the
self-avoiding walk (\textit{SAW}),  where the random walker is not permitted to
step on already visited sites, simulating the behavior of linear polymers.  
Although all investigated \textit{RW} models with memory  are similar in the
sense that they incorporate interactions between steps, they display quite
distinct asymptotic properties.  Therefore, they belong to universality classes
which are usually different from \textit{RW} or  \textit{SAW}, though they may
cross over to either \textit{RW} or \textit{SAW} behavior in some limits.
Common properties for describing the behavior of a walker are the exponents
$\nu$ and $k$, characterizing the scaling with time $t$ of the mean square
end-to-end distance 
\begin{eqnarray}
  \label{generell}
  \hspace*{1cm} \langle R^2(t)\rangle \sim t^{2 \nu} \nonumber \hspace*{6.5cm} (\arabic{equation}\text{a})\\
\text{and the average number of visited sites} \nonumber \hspace*{9.3cm} \\
   \hspace*{1cm} \langle S(t)\rangle \sim t^{k} .  \nonumber \hspace*{6.5cm} (\arabic{equation}\text{b})
\end{eqnarray}
\addtocounter{equation}{1}
For \textit{RW} the exponents are $\nu=1/2$ for all dimensions $d$, $k=1/2$ for
$d=1$ and $k=1$ for $d \ge 2$~\cite{RW}. For \textit{SAW}, $k=1$ for all $d$
and  $\nu \cong 3/(d+2)$~\cite{deGennes79,Barat95}. A comparative
study~\cite{Duxbury84} of some of these
models~\cite{Stanley83,Domb72,Amit83,Malakis75} in one dimension has shown that
the characteristic exponents depend crucially on the particular form of the
interaction between the steps.  Some of the important mechanisms are the range
of the interaction, the presence of cumulative memory effects, and the global
or local normalization conditions. Models with global and local normalization
conditions are also refered to in the literature as static and dynamic models,
respectively. 

Recently, one of the dynamic models without cumulative memory effects, the
self-attracting random walk (\textit{SATW})~\cite{Sapoz94}, has been found to
display,  in contrast to all other models, a swelling-collapse transition at a
critical attractive interaction $u_{\mathrm{c}}$ in $d \ge
2$~\cite{Ordemann/Berkolaiko/Bunde/Havlin:2000}. The characteristic exponents
$\nu$ and $k$ for \textit{SATW} are in different universality classes for
below  $u_{\mathrm{c}}$ (swollen walk), at $u_{\mathrm{c}}$, and above
$u_{\mathrm{c}}$ (collapsed walk). Above and below criticality, $\nu$ and $k$
have been determined analytically. At criticality, the exponents could only be
studied numerically,  and due to the finite-size effects close to a transition,
there have remained open questions regarding the asymptotic behavior of
\textit{SATW}. A careful analysis of the simulation data  and a scaling
approach different from the one developed
in~\cite{Ordemann/Berkolaiko/Bunde/Havlin:2000} is found to be necessary for a
comprehensive study of the structural properties of \textit{SATW}, especially
at critical $u=u_{\mathrm{c}}$, which is the focus of this paper. To determine
the fractal dimension of the cluster and the interface generated by the walker,
and to give more precise results for the characteristic exponents, we
investigate the temporal development of the number of sites visited by the
walker which have a certain amount of already visited next nearest neighbor
sites. Identifying the sites belonging to the external and internal perimeter
of the cluster in $d=2$, the fractal dimensions of these structures are studied
for attractions below, at, and above criticality. 

The paper is organized as follows: In Sec.~\ref{model}  several known
\textit{RW} models with interactions are summarized and their behavior is
compared. The \textit{SATW} model is briefly reviewed in Sec.~\ref{swell},
presenting the analytical and numerical results for the mean square end-to-end
distance and the average number of visited sites for varying strength of
attractive interaction  
following~\cite{Ordemann/Berkolaiko/Bunde/Havlin:2000}, showing the evidence
for the phase transition.  In Sec.~\ref{struc} we investigate the structure of
the cluster grown by the walker for different attractive interactions using a
new approach consisting of scaling arguments and Monte Carlo simulations which
leads to more insight for the behavior of the system at criticality. We  study
the fractal dimensions of the cluster and its interface in $d=2$ and $d=3$.
Closed form expressions for the average number of visited sites  and the  mean
time to visit a certain number of sites in $d=1$ are given in
Sec.~\ref{onedim}. Finally, in Sec.~\ref{concl} we summarize our main results.

\section{\textit{RW} models with interaction}\label{model}

When a self-avoiding constraint like in the \textit{SAW} model is added to the
\textit{RW}, the evolution of the walk becomes heavily dependent on the entire
history of the walk, converting it into a non-Markovian system. A bridge
between ordinary \textit{RW} and non-Markovian walks can be constructed by
associating energies $E_t(w)$ to all possible configurations $w$ of a $t$~step
random walk, defining an ensemble probability for a certain walk configuration
as $P_t(w) = \exp [ - \beta \, E_t(w)]/ \sum_w \exp [ - \beta \, E_t(w)]$, with
$\beta = (k_{\mathrm{B}} T)^{-1}$, where $k_{\mathrm{B}}$ is Boltzmann's
constant and $T$ is the absolute temperature. In the high-temperature limit all
walks become equally likely, but at finite temperatures the ensemble
probabilities of individual paths differ.  If $E_t(w)<0$ for walks with many
visited sites, walks prefering to explore new terrain dominate the system,
whereas if $E_t(w)>0$ for walks with many visited sites, the system is governed
by configurations of walks attracted to their own path. Models based on this
concept are by definition in the class of static models, for a comprehensive
overview see~\cite{RW}. One straightforward way of modelling attractive and
repulsive interaction between steps, known as the interacting
walk~\cite{Stanley83}, is to assign an energy $E_t(w)= g \, S_t(w)$ to a walk
configuration, where $S_t(w)$ is the number of visited sites of a $t$~step
\textit{RW} configuration. For interaction parameter $g=0$ the simple random
walk is recovered, while for $g<0$ the walk becomes repulsive, and the
characteristic exponents are $\nu=k=1$ in  $d=1$~\cite{Redner83,Chan84}. For
$g>0$ the walk is attractive and $k=1/3$ and $2/3$, in $d=1$ and $2$,
respectively, as well as $d/(d+2)$ in $d \ge 3$~\cite{Stanley83,Redner83,RW}. 

In a more generalized static model including cumulative memory
effects~\cite{Duxbury85} the energy of a walk configuration is $E_t(w,\alpha)=
g \, \sum_s n_t^{\alpha}(s)$, with $0 \le \alpha \le 2 $, where $n_t(s)$ is the
number of times a certain site $s$ has been visited after $t$ steps. For
cumulative memory parameter $\alpha=0$ the previous model of~\cite{Stanley83}
is recovered, and for $\alpha=2$ it is also known as the Domb-Joyce
model~\cite{Domb72,Domb83}, while for $g=0$ or $\alpha=1$ it reduces to simple
random walks. The model is repulsive for $\alpha < 1$ and $g<0$ or $\alpha> 1$
and $g>0$,  in the latter case Flory arguments give $\nu=(\alpha+1)/(2+\alpha
\, d -d)$ independent of $g$ for $d\le d_{\mathrm{c}}=2\, \alpha/(\alpha-1)$,
where $d_{\mathrm{c}}$ is the critical dimension. For the attractive regime
results are only known in $d=1$, where for $\alpha< 1$ and $g>0$ it exhibits
continuous varying exponents depending on $\alpha$ with 
$\nu=k=(1-\alpha)/(3-\alpha)$, while for the other attractive branch $ \alpha
>1 $ and $g<0$ it is always self-trapping  as $\langle R\rangle$ and $\langle
S\rangle$ saturate.

An approach analogous to the one above for the static models can be made for
the less investigated dynamic models, where the local normalization is done by
assigning a probability $P_i$ to the walker to step to  the next nearest
neighbor site $i$ \textit{during} the evolution of the walk,    with $  P_i =
\exp ( g \, n_i^{\alpha})/ \sum_{i=1}^{2 \, d} \exp ( g \, n_i^{\alpha})$,
$\alpha > 0$. Here, $n_i$ is the number of times the  neighbor site $i$ has
already been visited in the previous $t$ time steps~\cite{Oettinger85}. For
$g=0$ the simple random walk is recovered, while for $g<0$ the walk is
repulsive and the exponent $\nu$ is determined to be $\nu= (\alpha+1)/(2+\alpha
\, d)$ for  $d\le d_{\mathrm{c}}=2$. For the attractive regime $g>0$ it is
known that in $d=1$ the walk is always self-trapping. A special case of this
model is the true self-avoiding walk of~\cite{Amit83} with $g<0$ and
$\alpha=1$. 

In surprising contrast to the results for all above mentioned models, where the
characteristic exponents are always independent of the actual strength of the
attraction or repulsion parameter $g$, are the results for the  \textit{SATW}
model~\cite{Sapoz94} focused on here. In this model a random walker jumps with
probability $P \sim \exp(u \, n)$~\cite{negative} to a next nearest neighbor
site, with $n=1$ for already visited sites and $n=0$ for not visited sites. For
$u>0$, the walk is attracted to its own trajectory, so that  \textit{SATW} is
an extension of~\cite{Oettinger85} with attraction parameter $u=g>0$ in the
limit of no cumulative memory effect $\alpha \to 0$. Note that the results of
Ref.~\cite{Oettinger85} can not be directly applied to the \textit{SATW} model
as they are based on the restriction $\alpha >0$. Some representative examples
of two dimensional clusters grown  by \textit{SATW} for different values of 
$u$ at three distinct times $t$ of evolution are shown in Fig.~\ref{cluster}.
The exponents $k$ and $\nu$ of \textit{SATW} have been found to  depend on
$u$~\cite{Sapoz94,Sapoz98,Reis95,Lee98}, although it was not clear for some
time if $\nu$ and $k$ decrease continuously with increasing
$u$~\cite{Reis95,Lee98}, or if a critical value $u_{\mathrm{c}}$ exists, below,
at, and above which the exponents show different universal
behavior~\cite{Sapoz94,Sapoz98}. Recently, it was found by exhaustive computer
simulations up to $t=5 \cdot 10^9$ time steps that there exists a
swelling-collapse transition for \textit{SATW} at a critical attraction
$u_{\mathrm{c}}$~\cite{Ordemann/Berkolaiko/Bunde/Havlin:2000}, analogous to the
$\Theta$ transition in linear polymers at temperature $T=\Theta$ when an
attraction term $\exp(-A/T)$, $A<0$, is added to the self-avoiding
constraint~\cite{deGennes79,Barat95}. This phenomenon of a  swelling-collapse
transition can only occur because of a balance of the interaction energy and
the configurational entropy of the \textit{SATW} at criticality. When the
energy is not  of the order of the entropy as investigated
in~\cite{Oettinger85} for $\alpha > 0$,   the walk collapses for any attraction
$u=g>0$.

\section{Swelling-collapse transition for \textit{SATW}}\label{swell}

In~\cite{Ordemann/Berkolaiko/Bunde/Havlin:2000} it was shown that the
characteristic exponents $\nu$ and $k$ for \textit{SATW} are in different
universality classes for $u<u_{\mathrm{c}}$, $u=u_{\mathrm{c}}$, and
$u>u_{\mathrm{c}}$. Above $u_{\mathrm{c}}$, the walk collapses for $d \ge 2$,
and $\nu$ and $k$ are given by 
\begin{eqnarray}
  \label{exponents}
  \hspace*{1cm} \nu = \frac{1}{d + 1} \nonumber \hspace*{7cm} (\arabic{equation}\text{a})\\
\text{and} \nonumber \hspace*{15.8cm} \\
   \hspace*{1cm} k= \frac{d}{d + 1}  \; . \nonumber \hspace*{6.7cm} (\arabic{equation}\text{b})
\end{eqnarray}
\addtocounter{equation}{1}
Eqs.~(\ref{exponents}) follow
(cf.~\cite{Ordemann/Berkolaiko/Bunde/Havlin:2000}),  since for sufficiently
strong attraction $u > u_{\mathrm{c}}$ the grown clusters are compact
(see~Fig.\ref{cluster}), so the average number of visited sites scales with the
root mean square displacement $ \langle R\rangle \equiv {\langle
R^2\rangle}^{1/2} $ as $\langle S\rangle \sim {\langle R\rangle}^{d}$.
Comparing this to Eqs.~(\ref{generell}) yields $k = \nu d$ for $u >
u_{\mathrm{c}}$. Also, the mean cluster growth rate is proportional to the
ratio between the number of  boundary sites and the total number of the cluster
sites, ${\rm d} \langle S\rangle / {\rm d} t \sim {\langle R\rangle}^{d-1} /
{\langle R\rangle}^{d} \sim t^{-\nu}$~\cite{Sapoz98,Rammal83}. Thus $ \langle
S\rangle \sim t^{-\nu +1}$. Combining these results with Eq.~(\ref{generell}b),
one obtains  Eqs.~(\ref{exponents}). Below a critical interaction
$u_{\mathrm{c}}$, the walk swells and the exponents are as with no 
interaction~\cite{RW}, i.e. $\nu=1/2$ and $k=1$ for $d \ge 2$.  The above
analytic  arguments are in good agreement with numerical simulations in $d=2$
and $d=3$ (see Fig.~\ref{exponentsknu}). At $u_{\mathrm{c}}$, the exponents are
numerically determined to be  $\nu_{\mathrm{c}} = 0.40 \pm 0.01$ and
$k_{\mathrm{c}} = 0.80 \pm 0.01$ for $d=2$ as well as $\nu_{\mathrm{c}} = 0.32
\pm 0.01$ and $k_{\mathrm{c}} = 0.91 \pm 0.03 $  for
$d=3$~\cite{Ordemann/Berkolaiko/Bunde/Havlin:2000}. Note that for $d=1$ no
swelling-collapse transition exists, as the walk is collapsed for all $u$, and
Eqs.~(\ref{exponents}) reveals the known values $k=1/2$ and $\nu=1/2$ for
random walk  in $d=1$~\cite{RW,Prasad96} (see also Sec.~\ref{onedim}).

In the analogous static model of the interacting walk~\cite{Stanley83} with
attractive interaction and no cumulative memory effects, a phase transition can
not be observed  due to the fact that the global normalization condition
increases the weight of the interaction energy more than the local
normalization condition, as already observed in~\cite{Duxbury84} for $d=1$. 
This can also be seen from the asymptotic behavior of the characteristic
exponent $k=d/(d+2)$ for the attractive interacting walk~\cite{Stanley83,RW} in
comparison to the 'less collapsed' $k=d/(d+1)$, Eq.~(\ref{exponents}b), for the
\textit{SATW} discussed here. Therefore, a static interacting walk with the
slightest attraction $u=g>0$ has a qualitatively different behavior than the
ordinary \textit{RW} and is collapsed. Note that in both models $k$ never
becomes independent of the dimension, although it approaches unity from below
in the large dimensionality limit.

Due to finite time effects at criticality it is not possible to determine the
exponents $\nu_{\mathrm{c}}$ and $k_{\mathrm{c}}$  more and more accurately by
simply increasing the number $t$ of time steps performed by the walker.  As
long as the attraction is slightly above or below criticality, asit is always
the case in numerical simulations, the exponents will  finally cross over to
their expected values above resp.~below $u_{\mathrm{c}}$ after some time $t$.
Introducing a crossover time $t_\xi$, below which the exponent $\nu$ is close
to $\nu_{\mathrm{c}}$ and  above which $\nu$ approaches $1/2$ for $u <
u_{\mathrm{c}}$ and $1/(d+1)$ for $u > u_{\mathrm{c}}$ (see Fig.~1
of~\cite{Ordemann/Berkolaiko/Bunde/Havlin:2000}), the following scaling theory
holds: 
\begin{eqnarray}
  \label{scaling}
  \hspace*{1cm}  \langle R(t) \rangle \sim t^{\nu_{\mathrm{c}}} f_{\pm}(t/t_\xi) \; , \nonumber \hspace*{6.0cm} (\arabic{equation}\text{a})\\
\text{with} \nonumber \hspace*{15.4cm} \\
   \hspace*{1cm} t_\xi = |u-u_{\mathrm{c}}|^{-\alpha} \; , \nonumber \hspace*{6.4cm} (\arabic{equation}\text{b})
\end{eqnarray}
\addtocounter{equation}{1}
where the plus sign refers to $u > u_{\mathrm{c}}$, the minus sign to $u <
u_{\mathrm{c}}$, and the exponent $\alpha$~\cite{alpha} has to be determined 
numerically. As $t_\xi$ is the only relevant time scale, the scaling functions
bridge the short time and the long time regime. To match both regimes, it is
require that $f_{\pm}(x) = {\mathrm{const}}$ for $x \ll 1$ ($t \ll t_\xi$),
and  $f_{+}(x) \sim {x}^{1/(d+1) - \nu_{\mathrm{c}}}$,  $f_{-}(x) \sim {x}^{1/2
- \nu_{\mathrm{c}}}$ for $x \gg 1$. An analogous scaling approach holds for 
$\langle S(t) \rangle$, and an excellent data collapse can be obtained for
$\alpha = 7 \pm 1$ in $d=2$  and  $\alpha = 5.0 \pm 0.5$ in $d=3$ (see Fig.~3
of~\cite{Ordemann/Berkolaiko/Bunde/Havlin:2000}), confirming the numerical
values for $\nu_{\mathrm{c}}$, $k_{\mathrm{c}}$, and $u_{\mathrm{c}}$
determined from~Fig.~\ref{exponentsknu}.

Since the mass of the cluster generated by the walker, consisting of all
visited sites, scales as 
\begin{equation}
\langle  S  \rangle \sim {\langle R \rangle}^{k/\nu}, 
\label{df}
\end{equation}
the ratio $k/ \nu$ corresponds to the fractal dimension $d_{\mathrm{f}}$ of the
cluster,
\begin{equation}
d_{\mathrm{f}} = \frac{k}{\nu}. 
\label{knu}
\end{equation}
In $d=2$ the clusters are compact for all $u$ as $k/\nu = d_{\mathrm{f}}=d$. In
$d=3$ they are compact for $u > u_{\mathrm{c}}$, while for $u <
u_{\mathrm{c}}$, the fractal dimension of clusters generated by simple random
walks $d_{\mathrm{f}} =2<d $ is obtained.  At the criticality, $d_{\mathrm{f}}$
was found numerically to be $d_{\mathrm{f}}=2.84 \pm  0.25 $, but  the
possibility that $d_{\mathrm{f}} = d$ could not be ruled  out in
Ref.~\cite{Ordemann/Berkolaiko/Bunde/Havlin:2000}.

\section{The cluster and its interface at criticality}\label{struc}

To clarify if the grown clusters in $d=3$ at $u = u_{\mathrm{c}}$ are compact
or fractal, and to  learn more about the structure of the 
{\textit{SATW}}-clusters and their interfaces  at criticality, we consider the
following: Denoting by $N_i(t)$  the number of cluster sites which have $i$ of
their $2 \, d$ next nearest neighbor sites belonging to the cluster after $t$
time steps, then the number of all cluster sites $S(t)$ is the sum of all
$N_i(t)$
\begin{equation}
 S(t)=N_1(t) + N_2(t) + ... + N_{{2 \, d}}(t).
\label{sum}
\end{equation}
The cluster growth rate is {\it equal\/} to the probability to be on the
boundary of the cluster multiplied by the conditional probability to expand the
cluster while being on its boundary.  Suppose the walker is on a site which has
$i$ visited next nearest neighbor sites. As the probability to step to a next
nearest neighbor site  is $P \sim \exp(u \, n)$, with $n=1$ for already visited
sites and $n=0$ for   unvisited sites, the probability to jump on a visited
neighbor site is proportional to $i \,\exp(u)$, whereas the probability to jump
on an unvisited  neighbor site is proportional to $2 \, d-i$. Thus the
normalized probability $\tilde{P}_i$ to expand, i.e. to jump on one of the $2
\, d-i$ unvisited next nearest neighbor sites, is given by 
\begin{equation}
\label{P_i_def}
\tilde{P}_i = \frac{2 \, d-i}{ i\, \exp(u) + 2 \, d-i}.
\end{equation}
Therefore, the average cluster growth rate  is 
\begin{equation}
 \frac{{\rm d} \langle  S  \rangle}{{\rm d} t} = \tilde{P}_1 \, \frac{\langle  N_1  \rangle}{\langle 
 S  \rangle} + \tilde{P}_2 \,  \frac{\langle  N_2  \rangle}{\langle  S  \rangle} + ...  + \tilde{P}_{2 \,
 d-1}  \frac{\langle  N_{2 \, d-1}  \rangle}{\langle  S  \rangle}.
\label{elad1}
\end{equation}
The situation $i=2 \, d$ is special because $\tilde{P}_{2 \, d} =0$, as the
cluster cannot be expanded from a site where all surrounding sites already
belong to the cluster. Assuming that the average number of cluster sites
$\langle N_i \rangle$ which have $i$ already visited nearest neighbor sites
scales as
\begin{equation}
\langle N_i \rangle \sim {\langle S \rangle}^{a_i} \qquad \text{with} \qquad 0 \le a_i \le 1,
\label{elad2}
\end{equation}
there must be at least one $i $ for which $a_i=1$ to ensure that
Eq.~(\ref{sum}) holds.  In general, one can distinguish between two different
cases: 
\begin{itemize}
\item[(i)] there exists at least one $i < 2 \, d$ for which $a_i=1$
\item[(ii)] only $a_{2 \, d}=1$.
\end{itemize}
 In case (i), the average cluster  growth rate ${\rm d} \langle S \rangle /{\rm
d} t$ is dominated by  the $\langle N_i \rangle$ for which  $a_i=1$, leading to
${\rm d} \langle S \rangle / {\rm d} t \sim \tilde{P}_i \, \langle N_i \rangle
/\langle S \rangle = \tilde{P}_i \, {\langle S\rangle}^{a_i} /\langle S \rangle
=  {\rm const}$. Therefore, in this case one gets $k=1$ as $\langle S \rangle
\sim t$. In case (ii), when only $a_{2 \, d}=1$, the cluster  growth rate is
dominated by the $ \langle  N_i  \rangle$ for which $a_i =a_{\rm max}={\rm
max}(a_1, a_2,... ,  a_{2\, d-1})$. This gives ${\rm d} \langle S \rangle /{\rm
d} t \sim   {\langle S \rangle}^{a_{\rm max}} / \langle S \rangle = {\langle S
\rangle}^{a_{\rm max}-1}$, and hence
\begin{equation}
 k = \frac{1}{2 - a_{\rm max}} 
\label{alphamax}
\end{equation}
when compared to Eq.~(\ref{generell}b). As shown below, the grown cluster is 
fractal for case (i) while for case (ii) it is compact.

The above considerations also  enable us to determine numerically the exponent
$k$ in a way different  to Ref.~\cite{Ordemann/Berkolaiko/Bunde/Havlin:2000},
and additionally to gain more insight into the grown structures, especially in
the regime at  $u=u_{\mathrm{c}}$. Fig.~\ref{Dat} shows that the assumption of
Eq.~(\ref{elad2}) is clearly supported by numerical results. When plotting
$\langle N_i \rangle$ versus $\langle S \rangle$, the decision to which case
(i) or (ii) a certain  regime belongs becomes obvious from Fig.~\ref{Dat} for
all regimes except for  $u = u_{\mathrm{c}}$ in $d=3$. For  $u <
u_{\mathrm{c}}$ in $d=3$ we observe case (i) (see Fig.~\ref{Dat}d), as $a_i =1$
for all $i$. The other regimes belong to case (ii) (see 
Figs.~\ref{Dat}a,~\ref{Dat}b,~\ref{Dat}c, and~\ref{Dat}f), since only $a_{2 \,
d}=1$. At $u = u_{\mathrm{c}}$ in $d=3$ (Fig.~\ref{Dat}e), a more detailed
investigation of the results is necessary. By plotting the successive slopes
$a_i={\rm d}\ln \langle N_i \rangle/{\rm d}\ln \langle S \rangle$ of the data
of Fig.~\ref{Dat}e versus $1/ \ln \langle S \rangle$ (Fig.~\ref{Datsuc}a), it
becomes obvious that only $a_{2 \, d}=1$. Thus this regime is belonging to 
case (ii). Moreover, Fig.~\ref{Datsuc}a  indicates that all  $a_1, a_2,... , 
a_{2\, d-1}$ asymptotically have the same value, so $a_{\mathrm{max}} = a_1 =
a_2 = ... = a_{2\, d-1} $. As this is also found for all other regimes, in the
following we will denote $a_{\mathrm{I}} = a_i$ for all  $1 \le i \le 2 \, d-1
$ in both cases (i) and (ii), in distinction to $a_{2 \,d}$. The values of the
exponents are summarized in Table~\ref{tab2}, confirming the previous results
for the exponent $k$ (see Table~\ref{table}) when following
Eq.~(\ref{alphamax}).  At  $u = u_{\mathrm{c}}$ in $d=3$, $k_{\mathrm{c}}$ can
be determined more precisely than in
Ref.~\cite{Ordemann/Berkolaiko/Bunde/Havlin:2000} to be  $k_{\mathrm{c}} = 0.91
\pm 0.01$. Note that another approach to confirm our numerical results for $k$
is to plot $\langle N_i \rangle$ versus $t$ in a double logarithmic plot.
Denoting the resulting slopes as $\kappa_i$, we expect 
\begin{equation}
\kappa_i = k \, a_i
\label{kappa}
\end{equation}
as $\langle N_i \rangle \sim {\langle S \rangle}^{a_i} \sim t^{k \, a_i} \sim
t^{\kappa_i} $ when combining Eqs.~(\ref{generell}b) and~(\ref{elad2}). The
numerical values determined for $\kappa_i$ are summarized in Table~\ref{tab2}
and are in excellent agreement with the values determined for $a_i$ when
comparing with Eq.~(\ref{kappa}).

When examining the distinction between cases (i) and (ii), it becomes clear
that the  structures grown by the walker in case (i) are fractal, while the 
ones in case (ii) are compact. This can be explained by considering that in
case (ii) the sites which do not belong to the interface  dominate the growth
process as  $a_{2 \,d} > a_{\mathrm{I}}$, leading to a compact structure (i.e.
$d_{\mathrm{f}}=d$). In case  (i) all sites have the same contribution to the
growth of the structure due to  $a_{2 \,d} = a_{\mathrm{I}}$~\cite{Fig5}. As we
found that $u = u_{\mathrm{c}}$ in $d=3$ belongs to case (ii) (see
Fig.~\ref{Datsuc}a), the structure must be compact, and we can definitely
conclude that in this regime $d_{\mathrm{f}}$ has to be $d_{\mathrm{f}}=d=3$,
correcting the value $2.84 \pm 0.25$ obtained  earlier
~\cite{Ordemann/Berkolaiko/Bunde/Havlin:2000}. The latter value was calculated
by combining the numerical results for the exponents $k_{\mathrm{c}}=0.91 \pm
0.03 $ and $\nu_{\mathrm{c}}= 0.32 \pm 0.01$  following Eq.~(\ref{knu}). From
the new numerical result $k_{\mathrm{c}}=0.91 \pm 0.01$ and $\nu_{\mathrm{c}}=
k_{\mathrm{c}}/d_{\mathrm{f}}=k_{\mathrm{c}}/d$, we obtain the more accurate
estimate $\nu_{\mathrm{c}}=0.303 \pm 0.005$.

Next we focus on the interface of the grown cluster.  The total interface can
be divided into the external perimeter, which is usually the more interesting
fraction as it constitutes the reaction front with the environment, and the
internal perimeter, which is the boundary of the inner holes of the
cluster~\cite{addfootnote}. First we investigate the total interface. Its mass
$I$ is equal to  $N_1 + N_2 + ... + N_{{2 \, d-1}}$ and therefore scales as  
\begin{equation}
\langle I \rangle \sim {\langle S \rangle}^{a_{\mathrm{I}}}.
\label{interface}
\end{equation}
Combining Eqs.~(\ref{df}) and~(\ref{knu}) with Eq.~(\ref{interface}) yields
$\langle I \rangle \sim {\langle R \rangle}^{ a_{\mathrm{I}} \, d_{\mathrm{f}}
\,}$. Hence the  fractal dimension $d_{\mathrm{I}}$ of the interface is simply
\begin{equation}
d_{\mathrm{I}} = a_{\mathrm{I}} \, d_{\mathrm{f}}.
\label{dfI}
\end{equation}
The values for  $d_{\mathrm{I}}$ according to Eq.~(\ref{dfI}) are given in
Table~\ref{tab3}. For $u < u_{\mathrm{c}}$, we recover the  values for normal 
random walks~\cite{RW,citelogcorr}, $d_{\mathrm{I}}=2$ in $d=2$ and $3$. As the
fractal dimension of the external perimeter of a random walk is
$d_{\mathrm{EP}}=4/3 < d_{\mathrm{I}}$ in $d=2$~\cite{RW}, this suggests that
the internal perimeter is dominating the interface for  $u < u_{\mathrm{c}}$.
In contrast, in $d=3$  the external perimeter governs the  interface, 
$d_{\mathrm{EP}}=2 =  d_{\mathrm{I}}$~\cite{RW}, as three dimensional holes are
less likely than two dimensional ones due to geometrical constraints. For  $u >
u_{\mathrm{c}}$, we find $d_{\mathrm{I}}=d-1$, clearly confirming the
assumption that the structure of this regime is collapsed  (see
Fig.~\ref{cluster}). It forms a compact disk resp.~a compact sphere with a
rather smooth surface and holes  only in the surface layer. The clusters are
similar to the clusters grown in the Eden  model~\cite{Eden}, where each
unvisited next nearest neighbor site of the cluster has the {\it same} 
probability to be occupied at the given time step. It remains an open question
whether these clusters are in the same universality class, as one has to check
if the surface  for $u > u_{\mathrm{c}}$ resembles the  self-affine surface of
Eden cluster. At $u = u_{\mathrm{c}}$, the structure of the interface is
fractal in both $d=2$ and $3$ with the novel values of $d_{\mathrm{I}}=1.50 \pm
0.01$ and  $d_{\mathrm{I}}=2.73 \pm 0.03$, respectively. The rather surprising
result for $d_{\mathrm{I}}$ in $d=3$ can be understood when considering that
the structure of the interface  at $u = u_{\mathrm{c}}$ is more compact than
for $u < u_{\mathrm{c}}$  but the cluster
is not  collapsed as for $u > u_{\mathrm{c}}$. 

A straightforward way to confirm the above numerical results for
$d_{\mathrm{I}}$ is to  determine the fractal dimension of the interface   by
measuring it directly on the grown clusters instead of calculating it
dynamically during the growth process. Using the sandbox method~\cite{sandbox},
the outcome supports the results stated above, although they are less precise
due to the fact that the information about the growth process is lost. For the
results of $d_{\mathrm{I}}$ at criticality in $d=2$ and $d=3$ see
Fig.~\ref{interfacepic}, showing a good agreement with the values summarized in
Table~\ref{tab3} and  clearly excluding the value of $d_{\mathrm{f}}=2.84$
determined in~\cite{Ordemann/Berkolaiko/Bunde/Havlin:2000} for $d=3$, which
would lead to  $d_{\mathrm{I}}= 2.58$ when following Eq.~(\ref{dfI}).

A further interesting question is which perimeter, external or internal, is
dominating the interface of the clusters at criticality, and what is the actual
value of the non-dominating perimeter. Using the approach  of
Ref.~\cite{Gouyet:1987} to identify the sites $N_{\mathrm{EP}}$ belonging to
the  external perimeter of a cluster in $d=2$, and identifying the internal
perimeter sites $ N_{\mathrm{IP}}$ as the interface sites not belonging to
$N_{\mathrm{EP}}$, we expect  the average number of external perimeter sites $
\langle N_{\mathrm{EP}}  \rangle$ to scale as
\begin{eqnarray}
  \label{external}
  \hspace*{1cm} \langle N_{\mathrm{EP}} \rangle \sim {\langle S \rangle}^{a_{\mathrm{EP}}} , \nonumber \hspace*{6.5cm} (\arabic{equation}\text{a})\\
\text{and the average number of internal perimeter sites $ \langle
N_{\mathrm{IP}}  \rangle$ to scale as} \nonumber \hspace*{4.3cm} \\
   \hspace*{1cm} \langle N_{\mathrm{IP}} \rangle \sim {\langle S \rangle}^{a_{\mathrm{IP}}} .  \nonumber \hspace*{6.5cm} (\arabic{equation}\text{b})
\end{eqnarray}
\addtocounter{equation}{1}
Following the reasoning leading to Eq.~(\ref{dfI}) yields 
\begin{eqnarray}
  \label{dfEP}
  \hspace*{1cm} d_{\mathrm{EP}} = a_{\mathrm{EP}} \, d_{\mathrm{f}} \nonumber \hspace*{6.7cm} (\arabic{equation}\text{a})\\
\text{and} \nonumber \hspace*{15.8cm} \\
   \hspace*{1cm} d_{\mathrm{IP}} = a_{\mathrm{IP}} \, d_{\mathrm{f}} .  \nonumber \hspace*{6.7cm} (\arabic{equation}\text{b})
\end{eqnarray}
\addtocounter{equation}{1}
The results for $a_{\mathrm{EP}}$ (see Fig.~\ref{externalperimeter})  recover
the expected behavior of $d_{\mathrm{EP}}=4/3$ for $u< u_{\mathrm{c}}$, and
indicate that at criticality $d_{\mathrm{EP}}$ is clearly smaller than for 
below $u_{\mathrm{c}}$, although it is  difficult to determine the precise
value of $d_{\mathrm{EP}}$ from this plot. Thus, at $u_{\mathrm{c}}$ the
interface is dominated by the internal perimeter as
$d_{\mathrm{EP}}<d_{\mathrm{I}}$. This is confirmed by the  results for
$a_{\mathrm{IP}}$ at criticality, were the expected value of
$d_{\mathrm{IP}}=d_{\mathrm{I}}=1.5$ is recovered, although with
unsatisfactory  precision~\cite{running}.  Using the  sandbox
method~\cite{sandbox} to measure the fractal dimension of the  perimeter
directly, the results support  $d_{\mathrm{IP}}=1.5$ more precisely,  and we
find $d_{\mathrm{EP}}=1.25 \pm 0.05$ for $u=u_{\mathrm{c}}$  (see
Fig.~\ref{box}a). The fact that  $d_{\mathrm{EP}}$ for  $u < u_{\mathrm{c}}$ is
definitely smaller than  $d_{\mathrm{EP}}$ at  $u = u_{\mathrm{c}}$ reinforces 
that  at criticality the  \textit{SATW} model is in a new universality class. 
For  $u > u_{\mathrm{c}}$ one might expect that the interface is dominated by
the  external perimeter due to the collapsed structure of the cluster. However,
we find that the internal and the external perimeter contribute equally to the
interface, $d_{\mathrm{IP}}=d_{\mathrm{EP}}=1=d_{\mathrm{I}}$  (see
Fig.~\ref{box}b). 

In $d=3$, the fractal dimension of the external perimeter is   $d_{\mathrm{EP}}
= 2=d_{\mathrm{I}}$ for  $u < u_{\mathrm{c}}$,  supposing that
$d_{\mathrm{EP}}$ follows the random walk result~\cite{RW} like all other 
characteristic values determined for the \textit{SATW} in this regime. 
Unfortunately it is not possible to determine  $d_{\mathrm{EP}}$ or
$d_{\mathrm{IP}}$ in $d=3$ numerically using the algorithm
of~\cite{Gouyet:1987} to check  which one, the external or the internal
perimeter, governs the behavior at $ u_{\mathrm{c}}$, and to decide whether the
external perimeter dominates the interface for  $u>u_{\mathrm{c}}$,
i.e.~$d_{\mathrm{EP}}=d_{\mathrm{I}}> d_{\mathrm{IP}}$, or if
$d_{\mathrm{EP}}=d_{\mathrm{I}}= d_{\mathrm{IP}}$ in this regime. The values
for  $d_{\mathrm{EP}}$ and $d_{\mathrm{IP}}$ according to the above 
considerations are given in Table~\ref{tab3}.

\section{SATW in one dimension}\label{onedim}

As we mentioned above, in $d=1$ there is no swelling-collapse transition, as
the walk is collapsed for all $u$.  The exponents $k$ and $\nu$ are  $k=1/2$
and $\nu=1/2$~\cite{RW,Prasad96}, in accordance with Eqs.~(\ref{exponents}).
Following the considerations leading to Eq.~(\ref{alphamax}), the probability 
of the walker to be on the boundary of a \textit{SATW}-cluster is $\langle  N_1
\rangle / \langle S \rangle   = 2/ \langle S \rangle$, therefore \textit{SATW}
in $d=1$ belongs to case (ii), since $a_1=a_{\rm max}=a_{\rm I}=0$ and $a_{2 \,
d}=1$. Thus,  $k = 1/(2 - a_{\rm max})=1/2$ as expected from  analytical
results~\cite{Prasad96}. Note that, based on Eq.~(\ref{elad1}), one can derive
(although not  rigorously) a closed form expression for $\langle S(t) \rangle$
in $d=1$, extending the results of~\cite{Prasad96}. Since in $d=1$ the
conditional probability to expand the cluster while being on a perimeter site
is $\tilde{P}_1=1/(\exp(u) +1)$ (cf.~Eq.~(\ref{P_i_def})) and $N_1=2$, we
obtain
\begin{equation}
  \frac{{\rm d} \langle S(t) \rangle }{{\rm d} t} = \frac{1}{\exp(u)+1}\frac{2}{\langle S(t) \rangle } \, ,
 \label{1d}
\end{equation}
yielding
\begin{equation}
  \langle S(t) \rangle = \left(\frac{2t}{\exp(u)+1}\right)^{1/2} \, .
 \label{1dresult}
\end{equation}
This result is strongly supported by numerical simulations, see Fig.~\ref{1D}. 

In $d=1$ it is also possible to solve the inverse problem of deriving the
average time 
$\langle t(S) \rangle$ to visit a fixed number of visited sites $S$,
\begin{equation}
  \langle t(S) \rangle = S - 1 + \frac{(S-2) \, (S-1) \, (1-\tilde{P}_1)}{2 \, \tilde{P}_1}.
 \label{meantime}
\end{equation}
For asymptotically large $S$, Eq.~(\ref{meantime}) yields $ \langle t(S)
\rangle \sim S^2$, recovering the expected scaling with $k=1/2$. To derive
Eq.~(\ref{meantime}) we use the standard approach to \textit{{RW}} in $d=1$
(see, for instance, Ref.~\cite{Shiryaev}): The number of sites $S$ visited by a
walker in $d=1$ is equal to the span of the random walk.  The time $t(S)$
passed before the span reaches $S$ can be represented as the sum 
\begin{equation}
  t(S) = \sum_{i=2}^{S} \chi_i \, ,
  \label{app1}
\end{equation}
with $\chi_i = t(i) - t(i-1)$ being the  time spent before the span increases
from $i-1$ to $i$, given that the walker is initially at the boundary. Thus,
the mean time $\langle t(S) \rangle$ to  visit $S$ sites can be calculated as
\begin{equation}
\langle t(S) \rangle= \sum_{i=2}^{S} \langle \chi_i  \rangle \, ,
\label{app2}
\end{equation}
with $t(1)=0$ and $t(2)=1$.  Let us consider the nature of the variable
$\chi_i$ in some detail. When at the boundary, the walker  increases the span
with the probability of succeess $\tilde{P}_1$.  With probability
$1-\tilde{P}_1$ the walker stays inside the cluster and undertakes an excursion
until it hits the boundary again, which presents him with another opportunity
to increase the span.  We introduce the random variable $\mu$ which is the
number of unsuccessful attempts before the span is increased.  In other words,
$\mu$ is the number of excursions into the cluster, and  with probability
$\tilde{P}_1(1-\tilde{P}_1)^m$, $m\geq0$, $\mu$ is equal to $m$. Now we can 
decompose the  $\chi_i$ as
\begin{equation}
 \chi_i = 1+ \sum_{j=1}^{\mu} (\tau_j+1) \, ,
\label{app4}
\end{equation}
where $\tau_j$ is the lenght of the $j$-th excursion, and 1 is added to account
for the jump from the boundary into the cluster before the excursion started. 
The random variable $\tau_j$ can be viewed as the time a random walker started
at site 1 with the boundaries at 0 and $i-2$  needs to reach a boundary. Its
mean is known to be $\langle \tau_j \rangle = i-3$~\cite{Shiryaev}.  The mean
of $\chi_i$ is just the average number of excursions multiplied by the average
length of one excursion.  More rigorously, we express $\langle \chi_i \rangle $
in terms of conditional averages as follows
\begin{equation}
\langle \chi_i \rangle 
= \Bigg\langle 1 + \sum_{j=1}^{\mu} (\tau_j+1) \Bigg\rangle 
= 1 + \sum_{m=1}^{\infty}\, 
  \Bigg\langle \,\sum_{j=1}^{\mu} (\tau_j+1) \Bigg| \mu=m \Bigg\rangle\,
P(\mu=m)\,.
\label{app5}
\end{equation}
The variables $\tau_j$ are independent identically distributed random 
variables, therefore
\begin{equation}
\langle  \chi_i \rangle = 
1+ \sum_{m=1}^{\infty} m \,  \langle  \tau_j+1   \rangle \, P(\mu =m) =
1+ (i-2) \, \sum_{m=1}^{\infty} m \, \tilde{P}_1 \, {(1-\tilde{P}_1)}^m =
1+ \frac{(i-2)(1-\tilde{P}_1)}{\tilde{P}_1} \, .
\label{app6}
\end{equation}
By combining Eqs.~(\ref{app2}) and~(\ref{app6}) one obtains
Eq.~(\ref{meantime}).

\section{Concluding remarks}\label{concl}

In contrast to all other known random walk models with interaction,  the
\textit{SATW} model exhibits a swelling-collapse transition at a critical
attraction $u_{\rm c}$ in $d \ge 2$. The transition is similar to the
swelling-collapse transition observed at the `$\Theta$ point' $T=\Theta$ of
\textit{SAW} with an attraction term $\exp(-A/T)$, 
$A<0$~\cite{deGennes79,Barat95}. It can only arise because  the attractive
interaction energy  is of the same  order as the configurational entropy in $d
\ge 2$. Below $u_{\rm c}$ the entropy dominates and the walk is in the
universality class of simple \textit{RW}, above $u_{\rm c}$ the energy governs
the behavior and  the walk collapses. At $u_{\rm c}$, both contributions
balance each other, leading to a new universality class. In $d=1$, due to small
number of possible configuration caused by the geometrical constraints, the
walk is always collapsed, even without interaction $u=0$.

Analyzing the structure of the cluster grown by \textit{SATW} in detail, we
determined the fractal dimension of the cluster and its interface. In $d=2$,
the cluster is always compact while the interface has a fractal dimension
$d_{\mathrm{I}}=2$ and $1.50 \pm 0.01$ below and at $u_{\rm c}$, respectively,
dominated by the internal perimeter. Above $u_{\rm c}$, we found
$d_{\mathrm{I}}=d_{\mathrm{IP}}=d_{\mathrm{EP}}=1$. The fractal dimension of
the external perimeter at $u_{\rm c}$, $d_{\mathrm{EP}}=1.25 \pm 0.05$, is
smaller than below $u_{\rm c}$, $d_{\mathrm{EP}}=4/3$, demonstrating once more
the novel universality class at criticality. In $d=3$ the cluster is compact
for $u \le u_{\rm c}$, while for $u > u_{\rm c}$ the fractal dimension is
$d_{\mathrm{f}}=2$. Probably the most unexpected result is that in $d=3$ the
interface has a fractal dimension $d_{\mathrm{I}}=2$ above and below $u_{\rm
c}$, whereas at criticality it increases to $d_{\mathrm{I}}=2.73 \pm 0.03$.
This appealing structure at $u_{\rm c}$ is of interest on its own regarding
that many challenging problems in physics, chemistry, and biology are
associated with growth patterns in clusters and solidification fronts. The
results for the fractal dimensions of the perimeters could also be helpful when
investigating the structure of the cluster grown by \textit{SAW} at the
corresponding $\Theta$ transition, where sufficiently large clusters are
difficult to simulate because of the attrition of the walkers when stepping
into its own dead ends.

\section*{Acknowledgement}

Financial support from the German-Israeli Foundation (GIF), the
Minerva Center for Mesoscopics, Fractals, and Neural Networks and the Deutsche
Forschungsgemeinschaft is gratefully acknowledged.

\newpage

\begin{table}[b]
\begin{center}
\begin{tabular}{c c || c | c | c }
 &	& $u < u_{\mathrm{c}}$ & $u = u_{\mathrm{c}}$ & $u > u_{\mathrm{c}}$    \\\hline \hline
\multirow{3}{25mm}{\hspace{10mm} $d=2$}
 & $\nu$ & $1/2$ & $0.40 \pm 0.01$  & $1/3$   \\
 & $k $ & $1 $ & $0.80 \pm 0.01 $ & $ 2/3$  	  \\ 
 &   	& \multicolumn{3}{c}{$u_{\mathrm{c}}  =0.88 \pm 0.05$}  \\\hline
\multirow{3}{25mm}{\hspace{10mm} $d=3$}
 & $\nu $ & $1/2$ & $0.303 \pm 0.005$ & $1/4$  \\
 & $k $ & $1 $ & $ 0.91 \pm 0.01 $ & $ 3/4$  \\ 
 &  	& \multicolumn{3}{c}{$u_{\mathrm{c}}  =1.92 \pm 0.03$}  \\
\end{tabular}
\end{center}
\caption{The exponents $\nu$ and $k$ as well as  the estimated values for the
transition point $u_{\mathrm{c}}$ for {\textit{SATW}} in $d=2$ and $d=3$.}
\label{table} 
\end{table}

\begin{table}[b]
\begin{center}
\begin{tabular}{c c || c | c || c | c || c | c }
 & & \multicolumn{2}{c||}{$u < u_{\mathrm{c}}$} & \multicolumn{2}{c||}{$u = u_{\mathrm{c}}$} & \multicolumn{2}{c}{$u  > u_{\mathrm{c}}$} \\ 
 &	 & $a_i$ & $\kappa_i$ & $a_i$ & $\kappa_i$ & $a_i$ & $\kappa_i$  \\\hline \hline
\multirow{2}{25mm}{\hspace{10mm} $d=2$}
 & $1 \le i < 2 \,d $ & $0.95 \; (1) $ & $0.90 \; (1) $ & $0.75$ & $0.60$ & $ 0.50 $  & $ 0.33$ 	  \\ 
 & $i=2 \, d$ & $1$ & $0.95 \; (1)$  & $1$ & $0.80$ & $1$ & $0.67$  \\\hline
\multirow{2}{25mm}{\hspace{10mm} $d=3$}
 & $ 1 \le i < 2 \,d$ & $1 $ & $1$ &  $0.91$ & $ 0.83 $ & $0.67$ &$0.50 $ 		 \\ 
 & $i=2 \, d $ & $1$ & $1$   & $1$ & $0.91$ & $1$ & $0.75$   \\
\end{tabular}
\end{center}
\caption{The exponents $a_i$ and $\kappa_i$ for $1 \le i < 2 \,d$ and $i =2 \,
d $  for all three regimes $u < u_{\mathrm{c}}$, $u = u_{\mathrm{c}}$ and $u >
u_{\mathrm{c}}$ in $d=2$ and $d=3$ from numerical simulations. The errors are
of the order of one percent. For $u < u_{\mathrm{c}}$ in $d=2$, most probably
due to logarithmic corrections~\protect\cite{citelogcorr}, it is not possible
to obtain the asymptotic value of the exponents in numerical simulations,
therefore we give them in parentheses.  Note that the results for $a$ for the
regimes $u < u_{\mathrm{c}}$ and $u > u_{\mathrm{c}}$ are supported by
combining $k=1/2$ and Eq.~(\ref{alphamax}) as well as Eqs.~(\ref{exponents}b)
and~(\ref{alphamax}), respectively.}
\label{tab2} 
\end{table}

\begin{table}[b]
\begin{center}
\begin{tabular}{c c || c | c | c }
 &	& $u < u_{\mathrm{c}}$ & $u = u_{\mathrm{c}}$ & $u > u_{\mathrm{c}}$    \\\hline \hline
\multirow{4}{25mm}{\hspace{10mm} $d=2$}
 & $d_{\mathrm{f}}$ & $2$ & $2$  & $2$   \\
 & $d_{\mathrm{I}} $ & $2 $ & $1.50 \pm 0.01 $ & $ 1$  	  \\ 
 & $d_{\mathrm{EP}}$ & $4/3$ & $1.25 \pm 0.05$  & $1$   \\
 & $d_{\mathrm{IP}}$ & $2$ & $1.50 \pm 0.01$  & $1$   \\\hline
\multirow{2}{25mm}{\hspace{10mm} $d=3$}
 & $d_{\mathrm{f}} $ & $2$ & $3$ & $3$  \\
 & $d_{\mathrm{I}}$ & $2 $ & $2.73 \pm 0.03  $ & $ 2$  \\ 
\end{tabular}
\end{center}
\caption{The fractal dimension $d_{\mathrm{f}}$ of the cluster and
$d_{\mathrm{I}} = a_{\mathrm{I}} \, d_{\mathrm{f}}$ of the interface grown by
{\textit SATW} for all three regimes $u < u_{\mathrm{c}}$, $u =
u_{\mathrm{c}}$, and $u > u_{\mathrm{c}}$ in $d=2$ and $d=3$. Additionally in
$d=2$ the fractal dimensions of the external perimeter $d_{\mathrm{EP}}$ and 
the internal perimeter  $d_{\mathrm{IP}}$ are given. In $d=3$ it is 
particularly not clear whether  the external or the internal perimeter
dominates the interface at  $u = u_{\mathrm{c}}$. In case no errorbars are
given, the numerical results are supported by analytical considerations.}
\label{tab3} 
\end{table}

\newpage

\begin{figure}
\caption{Representative example of the structures grown by {\textit{SATW}}
after $t=2^{12}, 2^{14},$ and $2^{16}$ time steps for  $u =0 $, $u = 0.88 \,  (
\protect\cong u_{\mathrm{c}})$, and $u =2.5$ in $d=2$. Gray sites form the
interface, black sites are completely surrounded by already visited neighbor
sites. The length scale is arbitrary chosen so that the clusters fill the size
of the box. Note that for $u < u_{\mathrm{c}}$, one can  easily follow the
growth process because of the distinctive structure of the cluster, while for
$u = u_{\mathrm{c}}$ it is difficult to follow, since the walker keeps coming
back more often altering the structure. For $u > u_{\mathrm{c}}$ it is not
possible to do so as the grown clusters are compact.}
\label{cluster}
\end{figure}

\begin{figure}
\caption{The values of the exponents $k$~(squares) and $\nu$~(circles) vs
attraction $u$ in  $d=2$~(open symbols)  and  $d=3$~(filled symbols) for $t=
10^8$, obtained by a least square fit of the slope of $\ln \langle R^2(t)
\rangle$ and $\ln \langle S(t) \rangle $ vs $\ln t$ for large t, respectively. 
Note that for $u > u_{\mathrm{c}}$  the values of $k$ and $\nu$ approach the
theoretical predictions of Eqs.~(\ref{exponents}), marked as dashed lines. The
extimated values of $u_{\mathrm{c}}$ are $u_{\mathrm{c}} = 0.88 \pm 0.05$ in
$d=2$ and $u_{\mathrm{c}}= 1.92 \pm 0.03$ in $d=3$.}
\label{exponentsknu}
\end{figure}

\begin{figure}
\caption{The average number of cluster sites $\langle N_i(t) \rangle$ having
$i$ already visited nearest neighbor sites plotted vs the average number of all
cluster sites $\langle S(t) \rangle$ up to $t=2 \cdot 10^{9}$ time steps (for
$u<u_{\mathrm{c}}$ in $d=2$ and $d=3$ only up to $t= 10^{9}$ and $t=3 \cdot
10^{7}$, respectively) averaged over 100 configurations in $d=2$ for  {\bf (a)}
$u = 0.5 < u_{\mathrm{c}}$, {\bf (b)} $u = 0.88 \protect\cong u_{\mathrm{c}}$,
{\bf (c)} $u =2.5 > u_{\mathrm{c}}$, and in $d=3$ for  {\bf (d)} $u =1 <
u_{\mathrm{c}}$, {\bf (e)} $u =1.92   \protect\cong u_{\mathrm{c}}$, and {\bf
(f)} $u = 4 > u_{\mathrm{c}}$. In $d=2$ the data for the different values of
$i$ are marked by $i=1$~(circles), $i=2$~(squares), $i=3$~(diamonds), and 
$i=4$~(pluses), whereas in $d=3$,  $i=1$~(circles), $i=2$~(squares), 
$i=3$~(diamonds), $i=4$~(upward triangles), $i=5$~(downward triangles), and
$i=6$~(pluses). Note that  (d) clearly differs from (a),(b),(c), and~(f). The
value of the slopes $a_i$ determined from the data are summarised in
Table~\ref{tab2}.}
\label{Dat}
\end{figure}

\begin{figure}
\caption{{\bf (a)} The successive slopes $a_i = {\rm d}\ln \langle N_i
\rangle/{\rm d}\ln \langle S \rangle$ of the data from Fig.~\ref{Dat}(e)
plotted vs $1/ \ln \langle S \rangle$. A linear extrapolation of the points to
the limit $1/ \ln \langle S \rangle \rightarrow 0$ yields our estimates $a_i =
0.91 \pm 0.01$ for $1 \le i < 2 \,d $ and  $a_{2 \, d} = 1.00 \pm 0.01$,
clearly revealing that this regime belongs to  case (ii). The data is marked as
in Fig.~\ref{Dat}(e). {\bf (b)} The average number of cluster sites having $i$
already visited nearest neighbors divided by the average number of all cluster
sites $\langle N_i(t) \rangle / \langle S(t) \rangle$ vs $\langle S(t) \rangle$
of the data from Fig.~\ref{Dat}(a). Note that the data for $i=4$ marked by
pluses  shows a dominant behavior compared to the data for   $i=1$~(circles),
$i=2$~(squares), and $i=3$~(diamonds), implying that the regime  $u <
u_{\mathrm{c}}$ in $d=2$ belongs to case (ii).} 
\label{Datsuc}
\end{figure}

\begin{figure} 
\caption{The successive slopes $d_{\mathrm{I}} = {\rm d}\ln \langle I(R)
\rangle/{\rm d}\ln R$ of the mean mass $\langle I(R) \rangle$ of the interface
inside a disk resp.~a sphere of radius $R$ plotted vs $1/R$ for $u =
u_{\mathrm{c}}$ at $t=3\cdot 10^8$ resp.~$t=3\cdot 10^7$  time steps averaged
over 1000 configurations. A linear extrapolation of the points to the limit
$1/  R \rightarrow 0$ yields our estimates  {\bf (a)} $d_{\mathrm{I}}= 1.50 \pm
0.02$ in $d=2$ and {\bf (b)} $d_{\mathrm{I}}=2.72 \pm 0.04$ in $d=3$,
confirming the results summarized in Table~\ref{tab3}, which are marked by an
arrow. } 
\label{interfacepic}
\end{figure}

\begin{figure}  
\caption{The successive slopes $a_{\mathrm{EP}} = {\rm d}\ln \langle
N_{\mathrm{EP}} \rangle/{\rm d}\ln \langle S \rangle$  plotted vs $1/ \ln
\langle S \rangle$  averaged over 500 configurations for  $u =0.5<
u_{\mathrm{c}}$~(diamonds) up to $t=10^{7}$  time steps,  for $u =
u_{\mathrm{c}}$~(squares) up to $t=10^{8}$  time steps, and $u =2.5> 
u_{\mathrm{c}}$~(circles) up to $t=2\cdot10^{9}$ time steps in $d=2$ . A linear
extrapolation of the points to the limit $1/ \ln \langle S \rangle \rightarrow
0$ yields our estimates $a_{\mathrm{EP}} = 0.68 \pm 0.02$ and $ 0.50 \pm 0.02$
for below and above $u_{\mathrm{c}}$, respectively, confirming the results
summarized in Table~\ref{tab3}, which are marked by an arrow. At
$u_{\mathrm{c}}$, $a_{\mathrm{EP}}$ is clearly smaller than for below
$u_{\mathrm{c}}$.} 
\label{externalperimeter}
\end{figure}

\begin{figure}
\caption{  The successive slopes for $d_{\mathrm{EP}}= {\rm d}\ln \langle 
N_{\mathrm{EP}} \rangle/{\rm d}\ln R$  and $d_{\mathrm{IP}}= {\rm d}\ln
\langle  N_{\mathrm{IP}} \rangle/{\rm d}\ln R$ of the mean mass $\langle
N_{\mathrm{EP}}  \rangle$ and $\langle N_{\mathrm{IP}}  \rangle$ of the
external and internal perimeter, respectively, plotted vs $1/R$ for {\bf (a)}
$u = u_{\mathrm{c}}$ at $t=10^{8}$  time steps and {\bf (b)} $u =2.5 >
u_{\mathrm{c}}$ at $t=2\cdot10^{9}$ time steps averaged over 500 configurations
in $d=2$. A linear extrapolation of the points to the limit $1/  R \rightarrow
0$ yields our estimates  {\bf (a)}   $d_{\mathrm{IP}}=1.48 \pm 0.05$~(circles)
and {\bf (b)} $d_{\mathrm{EP}}=1.02 \pm 0.03$~(squares) and 
$d_{\mathrm{IP}}=1.01 \pm 0.03$~(circles), confirming the results summarized in
Table~\ref{tab3}, which are marked by an arrow. At $u_{\mathrm{c}}$,
$d_{\mathrm{EP}}$~(squares) is determined to be  $d_{\mathrm{EP}}=1.25 \pm
0.05$.}
\label{box}
\end{figure}

\begin{figure} 
\caption{The average number of visited sites $\langle S(t) \rangle$ in $d=1$
for different values of $u$, $u=2$~(circles), $u=4$~(squares),
$u=6$~(diamonds), and $u=9$~(triangles), plotted vs time $t$ scaled by
$\left(1+\exp(u)\right)^{-1}$ shows a good data collapse in agreement with
Eq.~(\ref{1dresult}) (solid line).  The plateau for small $t$ corresponds to
the average time (equal to $1+\exp(u)$) needed for the walker to escape the
initial cluster of the size 2.}
\label{1D}
\end{figure}


\begin{thebibliography}{MT1}
\bibitem{negative}
Note that in the original definition by Sapozhnikov $P \sim \exp (-nu)$ with
$u<0$~\cite{Sapoz98}, for convenience we use $P \sim \exp (+nu)$ with $u>0$.
\bibitem{Fig5}
Although  asymptotically $a=a_{2 \, d}=1$ in $d=2$ for $u < u_{\mathrm{c}}$
(see Table~\ref{tab2}), this regime belongs to case (ii) as $a_{2 \, d}$ is
dominating the growth as can be seen in Fig.~\ref{Datsuc}b where $\langle N_i
\rangle / \langle S \rangle$ is plotted versus $\langle S \rangle$, most 
probably due to logarithmic corrections of the
exponents~\protect\cite{citelogcorr}.
\bibitem{addfootnote}
A union $U$ of two fractal sets $A$ and $B$ with fractal dimension
$d_{\mathrm{A}}$ and $d_{\mathrm{B}}$, respectively,  has a fractal dimension
$d_{\mathrm{U}}={\mathrm{max}}(d_{\mathrm{A}},d_{\mathrm{B}})$~\cite{Eden}.
Here, the total interface is a union of the external and the internal
perimeter.
\bibitem{running}
The reason is insufficient data, since the running time for one walk 
configuration is considerably increased by the implementation of the algorithm
for the determination  of the external perimeter at several times $t$ during
the growth of the cluster.
\end{thebibliography}
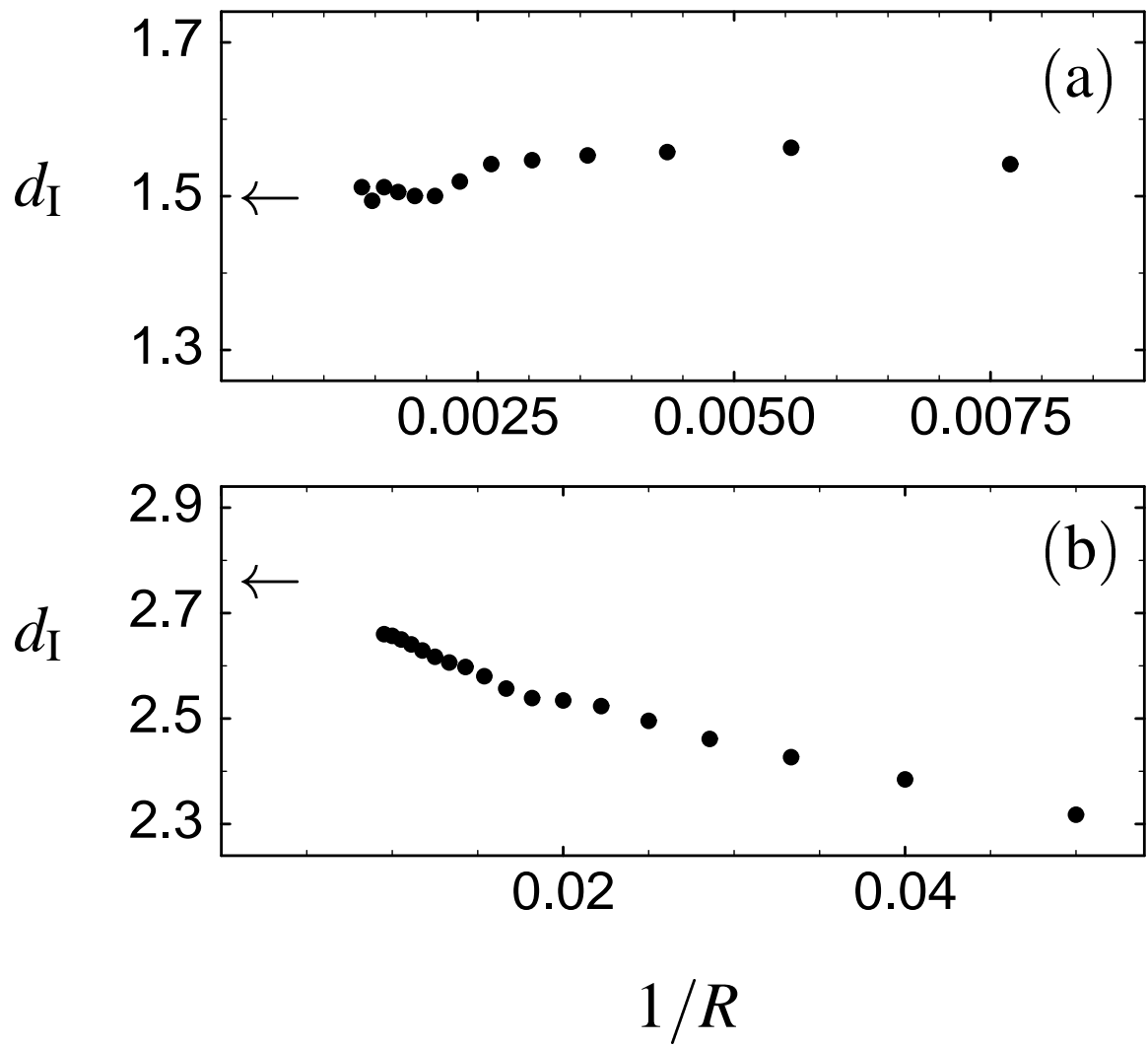
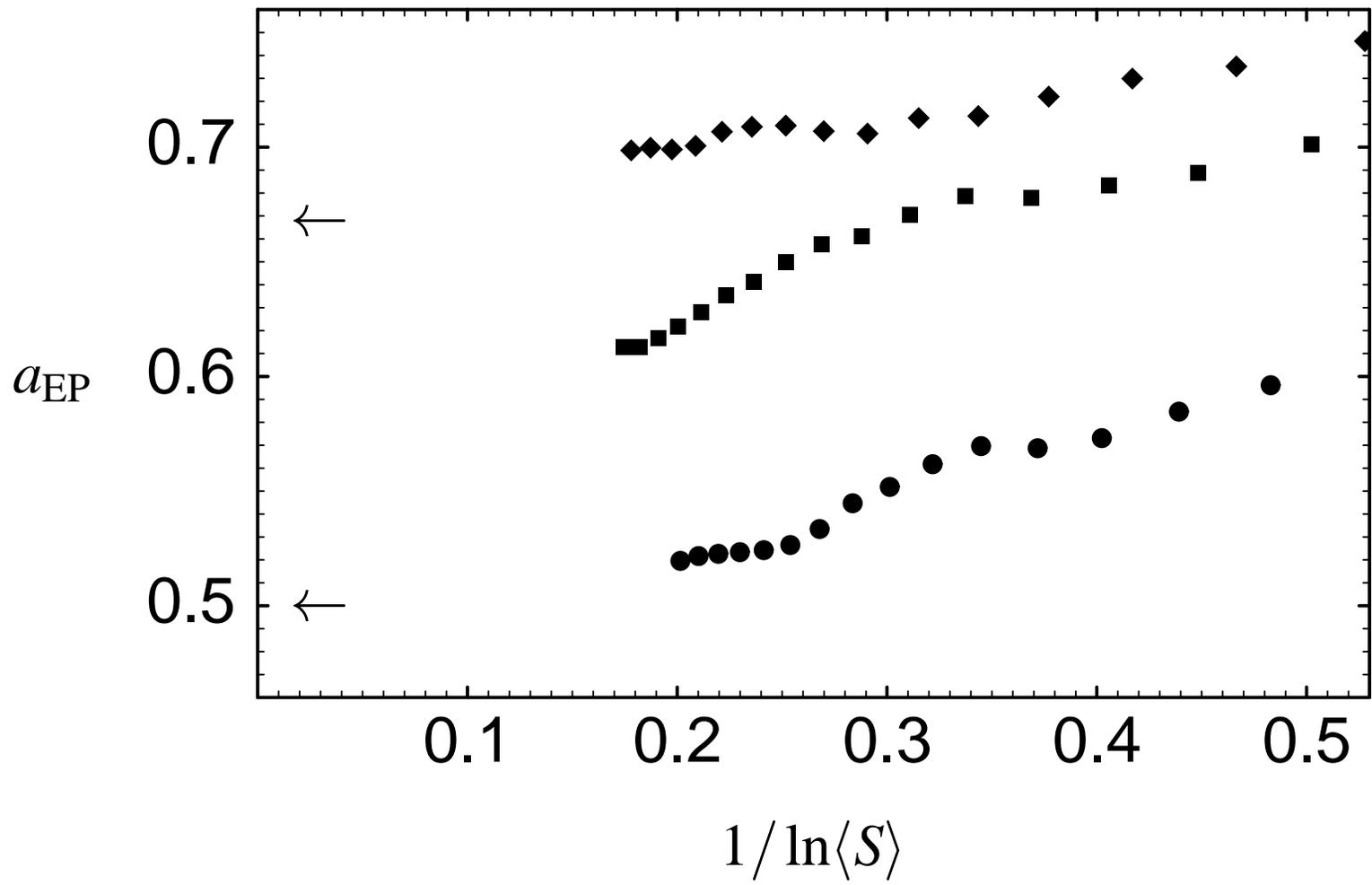
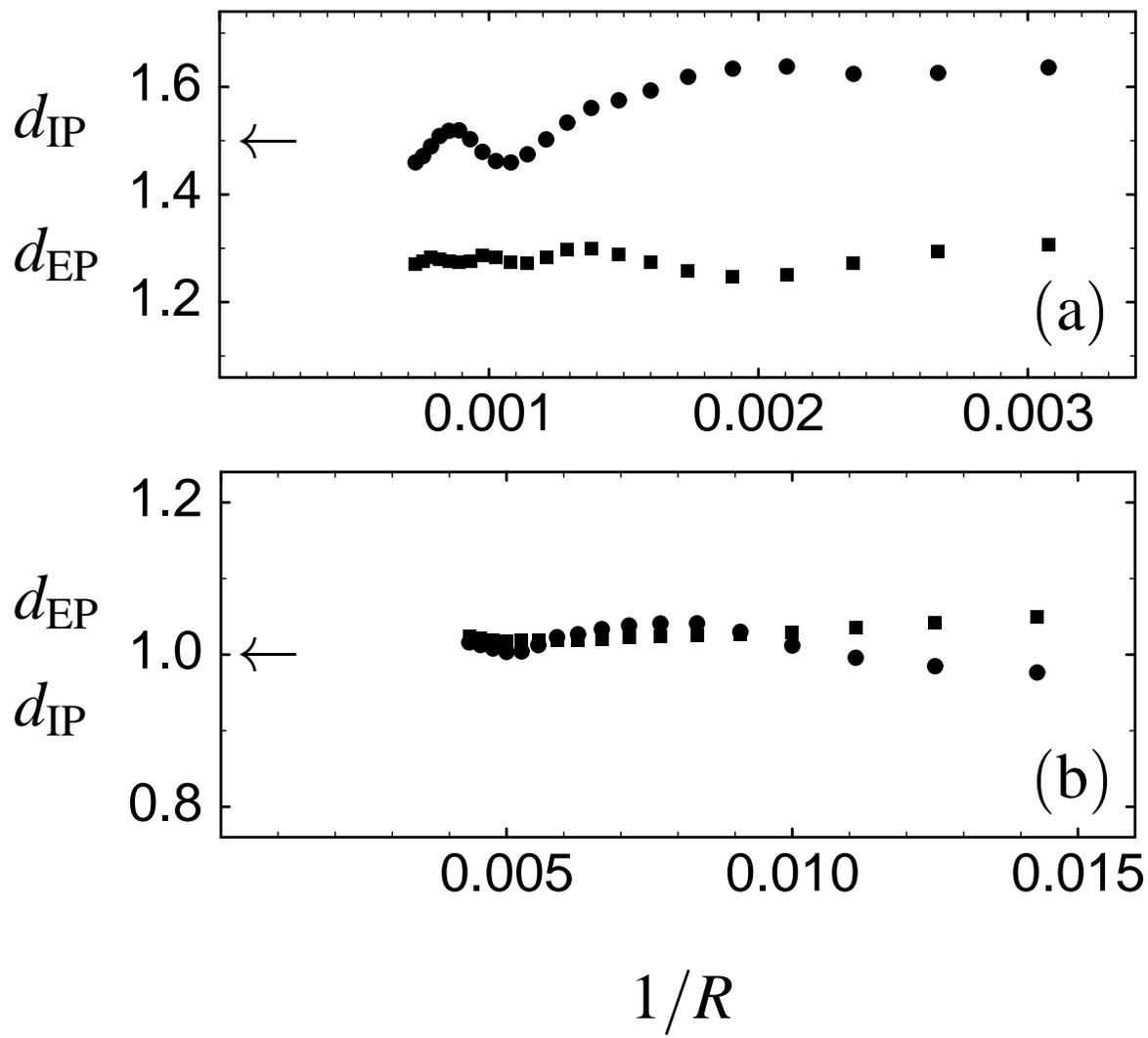
\end{document}